\documentclass[11pt,a4paper]{article}

\usepackage{geometry}
\usepackage[latin1]{inputenc}

\usepackage{comment}
\usepackage{float}
\usepackage{listings}
\usepackage{mathrsfs}
\usepackage{amsmath}
\usepackage{amssymb}
\usepackage{multicol}
\usepackage{float}
\usepackage{makeidx}
\usepackage{layout}
\usepackage{array}
\usepackage{hyperref}
\usepackage{latexsym}
\usepackage{indent first}

\usepackage{subcaption,graphicx}
\usepackage{appendix}
\usepackage{multirow}

\usepackage{color}

\usepackage[normalem]{ulem}

\newcommand{\RR}{\mathbb{R}}
\newcommand{\X}{\mathbf{X}}
\newcommand{\Y}{\mathbf{Y}}
\newcommand{\F}{\mathcal{F}}
\newcommand{\CC} {\mathcal{C}}
\renewcommand{\H}{\mathcal{H}}
\newcommand{\Phib}{\mathbf{\Phi}}
\newcommand{\Psib}{\mathbf{\Psi}}

\newcommand{\hs}{\hspace}

\newcommand{\ds}{\displaystyle}

\newcommand{\phib}{\overline{\phi}}
\newcommand{\psib}{\overline{\psi}}
\newcommand{\U}{\overrightarrow{U}}

\DeclareMathOperator{\sech}{sech}

\title{Pressure anomalies beneath solitary waves with constant vorticity}
\author{ Eduardo M. Castro$^{1}$, Marcelo V. Flamarion$^{2}$ and Roberto Ribeiro-Jr$^{3}$}

\date{}

\begin{document}

\maketitle

\begin{center}
{\footnotesize $^1$Graduate Program in Mathematics \\
   UFPR/Federal University of Paran{\' a} \\
   Centro Polit\'ecnico, Jardim das Am\'ericas, Curitiba-PR, Brazil, 81531-980\\
   eduardomdecastro@gmail.com
}\\
\vspace{0.3cm}
{\footnotesize $^2$Unidade Acad{\^ e}mica do Cabo de Santo Agostinho\\
	UFRPE/Rural Federal University of Pernambuco \\
	BR 101 Sul, Cabo de Santo Agostinho-PE, Brazil,  54503-900 \\
marcelo.flamarion@ufrpe.br }

\vspace{0.3cm}
{\footnotesize $^3$Departament of Mathematics\\
   UFPR/Federal University of Paran{\' a} \\
   Centro Polit\'ecnico, Jardim das Am\'ericas, Curitiba-PR, Brazil, 81531-980\\
   robertoribeiro@ufpr.br
}


\end{center}

	
	
	
	
\begin{abstract}

\noindent  While some works have investigated the particle trajectories and stagnation points beneath solitary waves with constant vorticity, little is known about the pressure beneath such waves. To address this gap,  we investigate numerically  the pressure beneath solitary waves in flows  with constant vorticity. Through a conformal mapping that flats the physical domain, we develop a numerical approach that allows to compute the pressure and the velocity field in the fluid domain. 
	 Our  experiments indicate that there exists a threshold vorticity such that 
	 pressure anomalies and stagnation points  occur when the intensity of the vorticity  is greater than this threshold. Above this threshold  the  pressure on the bottom boundary  has two points of local maxima and there are three stagnation points in the flow,  and below it the pressure has one local maximum and there is no stagnation point.
  \vspace{10pt}

\noindent {\bf Key words:} Constant vorticity, Solitary water waves, Euler equations, Pressure anomalies, Stagnation points.  
 
\end{abstract}
maketitle

\section{Introduction}

The study of water waves and its interactions with underline currents is a topic of research that  has piqued the curiosity of engineers,  mathematicians, physicists and oceanographers  over the centuries. Although many advances have already been achieved, there are a number of basic questions that are still open.

Currents are caused mainly by density differences in the water, tidal forces and by wind \cite{Dalrymple:1973}.  Mathematically,   wave-current interaction has been widely  investigated under the assumption  that the current is linearly sheared, i.e., flows with constant vorticity.  Physically, this can be representative of a realistic flow when waves are long compared with the depth or when waves are short compared with length scale of the vorticity distribution \cite{Teles:1988}.

 Flows with constant vorticity are mainly characterized by  the existence of overhanging  waves, the appearance of stagnation points and  the arise of pressure anomalies.
 
 Overhanging waves are free surface waves that are not graph of a function.  Among the numerical  studies in this direction it stands out the works of Vanden-Broeck \cite{Vanden-Broeck94,Vanden-Broeck96} in which the author finds  periodic and solitary overhanging waves and more recently the works of Dyachenko and Hur \cite{DyachenkoHur19a,DyachenkoHur19b}.  The existence of overhanging  waves is proved rigorously  by Constantin {\em et al. } \cite{ConstantinStraussVarvarica:2016} for periodic waves with constant vorticity, and more recently by Hur and Wheeler \cite{HurMiles:2022} for large or infinite depth. Although some theoretical  works have already allowed overhanging solitary wave profiles in their approach \cite{HaziotMiles:2021},  the rigorous proof of such type of  wave is still an open problem. 
 
 Stagnation points can be understood as fluid particles that are stationary in the wave moving frame. For irrotational flows it occurs at a sharp crest \cite{Varvaruca:2006} and in flows with constant vorticity they can emerge within the bulk of the fluid forming a recirculation zone whose profile resembles the Kelvin's cat's-eye flow.   The literature on stagnation points is extensive, starting with the work of Teles da Silva and Peregrine \cite{Teles:1988}, the reader is referred to the Ribeiro-Jr {\em et al.}
  to  a detailed study on the appearance of stagnation points beneath periodic waves with constant vorticity.  An overview of the works on stagnation point is given by Flamarion and Ribeiro-Jr \cite{Quarterly}. More recently, Ige and Kalisch \cite{IgeKalisch:2023} have investigated the particle trajectories associated with the propagation of periodic waves with constant vorticity in the framework of a new  Benjamin-Bona-Mahony equation.  
 
In irrotational flows the pressure exerted in the bulk of the fluid beneath a Stokes wave (a periodic travelling wave with  monotone profile from the crest to the trough) 
 attains its maximum on the bottom of the channel and below the crest. Besides, the pressure  is featured for being strictly increasing with the depth  and  is strictly decreasing horizontally away from a crest to a trough \cite{Constantin:2010}. Notable exceptions from these features arise in rotational flows with constant vorticity: (i) the maxima and minima of the pressure may occur within the bulk of the fluid; (ii) the pressure on the bottom can be out of phase with the surface elevation \cite{Teles:1988,Ali:2013,Vasan:2014, Strauss:2016, Ribeiro-Jr:2017}. The characteristics (i) and (ii) of the pressure are defined as pressure anomalies.  
 

 Although many advances have been accomplished on the understanding of the flow structure beneath waves with constant vorticity, it is unknown whether the  pressure anomalies known for periodic waves with constant vorticity also occur for solitary waves. Strauss and Wheeler \cite{Strauss:2016} have proved that overhanging  periodic or solitary waves must have a pressure sink, i.e., the pressure achieves its minimum within the bulk of the fluid and not on the free surface.  However, this is still an open question for free surface waves that are  graph of a function.  
  This issue was raised recently by Kozlov {\em et al.} \cite{Kozlov: 2020}.  In their  words, the following question  was raised: {`` Is the pressure  beneath a solitary wave in a flow with constant vorticity different from
 	the one in the irrotational case?''}
 
 In this work, we address  the question above.  
  The novelty is twofold: (i) we find numerically that,  when the vorticity  crosses a threshold,   the pressure  on the bottom boundary caused by a  solitary wave on the free surface can have    two points of local maxima; (ii) we analyse in the details the appearance of stagnation points beneath solitary waves. Thus, the paper at hand responds the question raised by  Kozlov {\em et al.} \cite{Kozlov: 2020}. Besides, it is a complement of  the studies  carried out by  Vasan and Oliveras \cite{Vasan:2014} and   Ribeiro-Jr {\em et al. }\cite{Ribeiro-Jr:2017} who have showed numerically the occurrence of pressure anomalies beneath periodic waves with constant vorticity  and analysed the appearance of stagnation points beneath such waves.  The approach used to compute the pressure and the stagnation points  consists in   determining  a conformal mapping under which   the physical domain is the image of a strip (canonical domain), then all calculations are made through pseudo-spectral methods.

   In summary, the results presented in this work are of interest of theorists and experimentalists. For a theorist, it can provide physical insights to a rigorous proof of the pressure anomalies. Likewise,  
  it  may inspire more  experimental studies on this topic since the phenomenon can be observed manipulating the intensity of the underlying current.

   For reference, this article is organized  as follows.   The governing equations of water waves in flows with constant vorticity are presented  in   section \ref{Sec_governing_eq}. In section \ref{Sec_steady}, we describe the conformal mapping and  the numerical method.  Then, we  
   present the  results in  section \ref{Sec_Results} and proceed to our  final considerations.
   

\section{Governing equations}\label{Sec_governing_eq}

We consider an incompressible flow of an inviscid fluid with constant density ($\rho$) in a  two-dimensional channel with finite depth ($d$) under the force of gravity ($g$).  Besides, we assume that the flow  is in the presence of a linearly sheared current (constant vorticity). Denoting the velocity field in the bulk of the fluid by $\U(x,y,t)=(u(x,y,t),v(x,y,t))$ and the free surface by ${\zeta}(x, t)$, this free-boundary problem can be described by the Euler equations
\begin{align}
\label{eq:euler1}
	&	\U_t+(\U \cdot \nabla ) \U  =  -\ds\frac{\nabla p}{\rho} -g\mathbf{j} \quad  \text{in} \; -d < y < \zeta(x,t),  \\
		&\nabla \cdot \U  =  0  \quad \text{in} \; -d < y < \zeta(x,t), \\
		&p  =  P_{atm} \quad  \text{at} \; y = \zeta(x,t), \\
		&v  =  \zeta_t+u\zeta_x  \quad \text{at} \; y = \zeta(x,t), \\
		\label{eq:euler5}
		&v  =  0 \quad  \text{at}  \; y=-d,		
	\end{align}
where $\mathbf{j}$ is the unitary vector $(0,1)$ and $P_{atm}$ is the atmospheric pressure. 

The assumption of constant vorticity enables us to write the velocity field as
	\begin{equation}\label{Ucampo}
		\U=\overrightarrow{U_0}+\nabla \phib,
	\end{equation}
	where 
	$$\overrightarrow{U_0}=(a y+f,0),  \quad f \in \RR,$$
	is a  linear shear flow solution of (\ref{eq:euler1})--(\ref{eq:euler5})  characterized by the flat surface $\zeta \equiv 0$ and constant vorticity $-a$. Here,   $\phib$ is the velocity  potential of an irrotational perturbation of the shear flow.

Equations (\ref{eq:euler1})--(\ref{eq:euler5}) are written in terms of $\phib$, then   non-dimensionaled  via the  transformation 		(\ref{rescale})
	\begin{equation}
	\label{rescale}
		\begin{array}{lllcll}
			x=dx', && \zeta=d\zeta', && \Omega=\ds\frac{ad}{\sqrt{dg}}, \\ 
			y=dy', && \phib=d\sqrt{dg}\phib', & & p=P_0+\rho gdp', \\
			t=\sqrt{\frac{d}{g}}t', && \psib=d\sqrt{dg}\psib', && F=\ds\frac{f}{\sqrt{dg}}.
		\end{array}
	\end{equation}
	Dropping the prime notation, this gives us the dimensionless version of  the governing equations	
	\begin{align}
	\label{eq:potential1}
		&			\Delta \phib = 0  \; \text{in} \; -1 < y < \zeta(x,t),  \\
		&	\zeta_t + (\Omega\zeta+F+\phib_x)\zeta_x=\phib_y  \: \text{at} \; y = \zeta(x,t), \\
		&	\phib_t+\frac{1}{2}(\phib_x^2+\phib_y^2)+(\Omega\zeta+F)\phib_x+\zeta-\Omega\psib=B(t) \; \text{at} \; y=\zeta(x,t), \\
			& \phib_y = 0  \; \text{at} \; y=-1,
		\end{align}
	where $-\Omega$ is the dimensionless vorticity,  $F$ is the Froude number and the pressure in fluid body is given by
	\begin{equation}\label{eq:pressure}
	p = -\left(\phib_t+\frac{1}{2}(\phib_x^2+\phib_y^2)+(a\zeta+f)\phib_x+\zeta-a\psib - B(t)   \right).
	\end{equation}

	For the study of traveling wave solutions it is convenient to eliminate time from the problem by passing to a  moving frame
	$$X=x-ct \quad \mbox{ and } \quad Y=y,$$
	where $c$ is the wave speed,  to be determined {\it a posteriori}.   In this new moving reference frame the wave is stationary and the flow is steady. Taking this new frame of reference into account, equation (\ref{eq:potential1})--(\ref{eq:pressure}) are written as
	\begin{align}
\label{eq:travelling1}
	&	\Delta \phib = 0  \quad \text{in} \; -1 < Y < \zeta(X),  \\
	\label{eq:kinematic}
	&	-c\zeta_X + (F+\Omega\zeta+\phib_X)\zeta_X=\phib_Y \quad \text{at} \; Y = \zeta(X), \\
		\label{eq:bernoulli}
	&	-c\phib_X+\frac{1}{2}(\phib_X^2+\phib_Y^2)+(\Omega\zeta+F)\phib_X+\zeta-\Omega\psib=B \quad \text{at} \; Y=\zeta(X), \\
	\label{eq:travelling2}
	&	\phib_Y = 0  \quad \text{at} \; Y=-1,
\end{align}
and
	\begin{equation}\label{eq:pressure2}
	p = -\left(-c\phib_X+\frac{1}{2}(\phib_x^2+\phib_y^2)+(a\zeta+f)\phib_x+\zeta-a\psib - B   \right).
\end{equation}

	We assume that $\zeta(X)$ is a solitary wave whose crest is located at $X=0$ and satisfies
\begin{equation}\label{limite}
	\zeta(X) \to 0 \quad \mbox{as} \quad \vert X \vert \to \infty. 
\end{equation}
In the following, we present a numerical scheme to compute the solutions of the system (\ref{eq:travelling1})--(\ref{eq:travelling2})   and to calculate the pressure in the fluid body via formula (\ref{eq:pressure2}).

\section{The conformal mapping  and the numerical method } \label{Sec_steady}

Since $\zeta(X)$ decays to zero as $\vert X \vert \to \infty $, we can truncate its infinite domain  to a finite one $[-\lambda/2,\lambda/2]$ with $\lambda >0$, and approximate the boundary conditions by periodic conditions. Then we can solve equations (\ref{eq:travelling1})--(\ref{eq:travelling2})   through  the conformal mapping technique  introduced by Dyachenko et al. \cite{Dyachenko},  that has been widely applied in  free boundary problems \cite{Choi, Milewski,Ribeiro-Jr:2017}.  	  This strategy consists in using a  conformal mapping  from a strip of length $L$ and width $D$ (canonical domain) onto the flow domain of the solitary wave $\{(X,Y)\in \RR^2, -\lambda/2\leq X \leq \lambda/2 \mbox{ and } -1\leq Y \leq \zeta(X)  \}$.  This map is such that  in the canonical domain the free boundary problem  (\ref{eq:travelling1})--(\ref{eq:travelling2})  can be solved numerically by the use of a spectral collocation method and Newton's method. 

\subsection{Conformal mapping}

\begin{figure}[!h]
	\centerline{\includegraphics{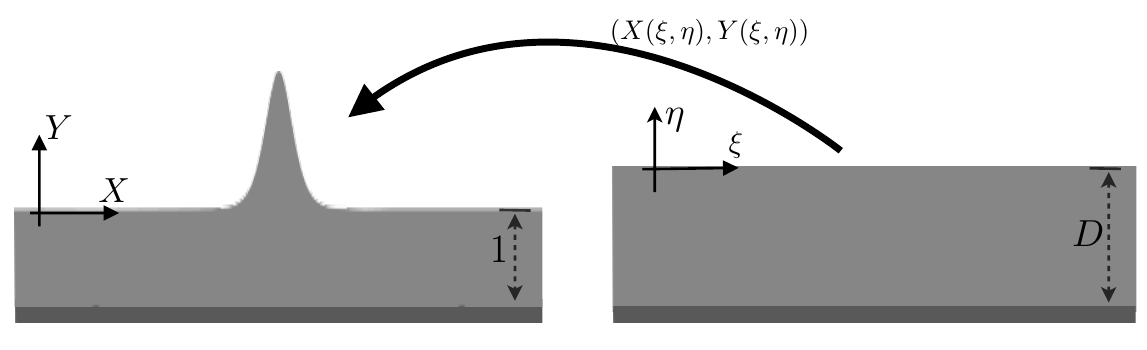}}
	\caption{Illustrative depiction of the conformal mapping. The free surface is flattened out in the canonical domain. }
	\label{Map}
\end{figure}

Consider the conformal mapping 
\begin{equation}
	Z(\xi,\eta)=X(\xi,\eta)+iY(\xi,\eta),
\end{equation}
under which the strip $\{(\xi,\eta)\in \mathbb{R}^2; \;  -L/2 \leq \xi \leq L/2 \mbox{ and } -D\leq \eta \leq 0\}$ is mapped onto the flow domain, as in Figure \ref{Map}. The constant $D$ will be determined so that  the canonical and the physical domain have the same length.   Since $Z$ is taken to be conformal, thus analytical, $X$ and $Y$ are actually conjugate harmonic functions, whereas the mapping's Jacobian is given by

\begin{equation}
	J=X_\xi^2+Y_\xi^2.
\end{equation}

A central characteristic of this mapping is given by the way the boundary curves from each domain are related

\begin{equation}\label{waveprofile}
	\begin{cases}
		Y(\xi,0)=\zeta(X(\xi,0)), \\
		Y(\xi,-D)=-1,
	\end{cases}
\end{equation}
which serves as Dirichlet data for the Laplace equation for $Y(\xi,\eta)$.  By denoting $\mathbf{Y}(\xi)=Y(\xi,0)$ and $\mathbf{X}(\xi)=X(\xi,0)$ the traces of the respective harmonic functions along $\eta = 0$, we have that

\begin{equation}\label{Ything}
	Y(\xi,\eta)=\mathcal{F}^{-1}\left[\frac{\sinh(k(\eta+D))}{\sinh(kD)} \mathcal{F}(\mathbf{Y})\right]+\frac{(\eta+D)\langle\Y\rangle + \eta}{D}, \quad k\neq 0,
\end{equation}
where $k=k(j)=(\pi/L)j$, for $j \in \mathbb{Z}$,  $\mathcal{F}$ is the Fourier transform in $\xi$-variable given by 

\begin{equation*}
	\mathcal{F}(f(\xi))=\hat{f}(k)=\ds\frac{1}{L}\ds\int_{-L/2}^{L/2} f(\xi)e^{-ik \xi} d\xi,
\end{equation*}

\begin{equation*}
	\mathcal{F}^{-1}(\hat{f}(k))=f(\xi)=\ds\sum_{j\in \mathbb{Z}} \hat{f}(k) e^{ik\xi}, 
\end{equation*}
and $\langle \,\cdot \, \rangle$  denotes the average defined by
\begin{equation*}
	\langle\Y\rangle=\ds\frac{1}{L}\int_{-L/2}^{L/2} \Y(\xi) d\xi.
\end{equation*}
 Differentiating   equation (\ref{Ything}) with respect to $\eta$ and  integrating   the Cauchy-Riemann equation $X_\xi=Y_\eta$, we get
\begin{equation} \label{Xeq}
	X(\xi,\eta)= \left(\frac{1+\langle\Y\rangle}{D}\right)\xi - \mathcal{F}^{-1}\left[\frac{i\cosh(k(\eta+D))}{\sinh(kD)}  \mathcal{F}(\mathbf{\Y})\right] , \quad k \neq 0.
\end{equation}

The canonical depth $D$ can be fixed if we require that both canonical and physical  domains have the same length. Let $L$ and $\lambda$ be the respective lengths, thus
$$\X(\xi = L/2) - \X(\xi = -L/2)= \lambda. $$
It follows from (\ref{Xeq}) that  this  restriction  leads to the relation
\begin{equation}\label{depth}
	D=1+\langle\Y\rangle.
\end{equation}
A reader interested in further details of the conformal mapping presented here should consult Flamarion and Ribeiro-Jr \cite{IJNumFluid} to this conformal mapping in the context of uneven topographies and its accuracy.

The Laplace equation is conformally invariant. So, denoting by  $\phi(\xi,\eta)=\overline{\phi}(X(\xi,\eta),Y(\xi,\eta))$ and   $\psi(\xi,\eta)=\overline{\psi}(X(\xi,\eta),Y(\xi,\eta))$ the potential and its harmonic conjugate in the canonical coordinates, one can easily obtain that:

\begin{equation*}
	\begin{array}{ll}
		\phi_{\xi\xi}+\phi_{\eta\eta}=0  & \hspace{10mm} \text{in} \; -D<\eta<0,\\
		\phi=\mathbf{\Phi}(\xi) & \hspace{10mm} \text{at}  \; \eta=0, \\
		\phi_\eta=0 & \hspace{10mm} \text{at} \; \eta=-D,
	\end{array}
\end{equation*}
and
\begin{equation*}
	\begin{array}{ll}
		\psi_{\xi\xi}+\psi_{\eta\eta}=0  & \hspace{10mm} \text{in} \; -D<\eta<0,\\
		\psi=\mathbf{\Psi}(\xi) & \hspace{10mm} \text{at}  \; \eta=0, \\
		\psi=Q & \hspace{10mm} \text{at} \; \eta=-D,
	\end{array}
\end{equation*}
where $Q$ is an arbitrary constant. The formulas for $\phi(\xi,\eta)$ and  $\psi(\xi,\eta)$ can be found in similar fashion to that worked out to $X(\xi,\eta)$ and  $Y(\xi,\eta)$, which yields 
\begin{equation*}
	\phi(\xi,\eta)=\mathcal{F}^{-1}\left[\frac{\cosh(k(\eta+D))}{\cosh(kD)} \mathcal{F}(\mathbf{\mathbf{\Phi}})\right], 
\end{equation*}

\begin{equation*}
	\psi(\xi,\eta)=\mathcal{F}^{-1}\left[\frac{\sinh(k(\eta+D))}{\sinh(kD)} \mathcal{F}(\mathbf{\mathbf{\Psi}})\right] - Q \frac{\eta}{D}.
\end{equation*}
Using the Cauchy-Riemman equation $\phi_\xi = \psi_\eta$ and evaluating along $\eta = 0$ we find that 

\begin{equation}\label{Phimap}
	\mathbf{\Phi}_\xi(\xi)=  \mathcal{F}^{-1}\left[ -i\coth(kD) \mathcal{F}_k(\mathbf{\Psi}_\xi)\right].
\end{equation}

For simplicity, we make use of the Fourier operator $\CC[\cdot]$ defined as follows: given a function $h(\xi)$, 
\begin{equation}
	\CC[h(\xi)]=\CC_0[h(\xi)]+\lim_{k \to 0} i\coth(kD)\hat{h}(k),
\end{equation}
where $\CC_0[\;\cdot\,]=\F^{-1}\H\F[\;\cdot\,]$, with  $\H$ given by

\begin{equation}
	\H (k)=\begin{cases}
		i \coth(kD), \; k \neq 0 \\
		0, \; k=0.
	\end{cases}
\end{equation}
For the particular case of $\CC[\cdot]$ evaluated at $h_\xi(\xi)$, we have that
\begin{equation}
	\CC[f_\xi(\xi)]=\CC_0[h_\xi(\xi)] - \frac{\hat{h}(0)}{D},
\end{equation}
With this notation, we obtain  from relations (\ref{Xeq}), (\ref{depth}) and (\ref{Phimap}) that 
\begin{equation}\label{pairs}
	\mathbf{X}_\xi=\ds 1-\CC_0[\mathbf{Y}_\xi] \\
\end{equation}
\begin{equation}\label{Phi_xi1}
	\mathbf{\Phi}_\xi=-\CC_0[\mathbf{\Psi}_\xi] +\dfrac{\hat{	\mathbf{\Psi}}(0)}{D}.
\end{equation}

Performing straight-forward calculations  we obtain that the Kinematic condition (\ref{eq:kinematic}) and Bernoulli law  (\ref{eq:bernoulli})  in canonical coordinates are given by

\begin{equation}\label{Psi}
	\Psib_\xi=c\Y_\xi-(\Omega \Y + F)\Y_\xi,
\end{equation}

\begin{equation}\label{ber2}
	-c\ds\frac{\Phib_\xi \X_\xi +\Psib_\xi \Y_\xi}{J}+\ds\frac{1}{2J}(\Phib_\xi^2+\Psib_\xi^2)+\Y+(\Omega \Y + F)\ds\frac{\Phib_\xi \X_\xi +\Psib_\xi \Y_\xi}{J}-\Omega \Psib = 0.
\end{equation}
Then, integrating  (\ref{Psi})  we get
\begin{equation}\label{Psi2}
	\Psib=c\Y-\left(\frac{\Omega \Y^2}{2} + F\Y\right) + M,
\end{equation}
where $M$ is an arbitrary  constant.  In order to simplify the use of the formula  (\ref{Phi_xi1})  we choose $\Psib$ so that $\hat{\mathbf{\Psi}}(0)=0$. This leads naturally to 
$$ M = \left< c\Y-\left(\frac{\Omega \Y^2}{2} + F\Y\right)\right>.  $$ 
Hence, in which follows 
\begin{equation}\label{Phi_xi}
	\mathbf{\Phi}_\xi=-\CC_0[\mathbf{\Psi}_\xi].
\end{equation}

By substituting equation (\ref{Psi2}) and (\ref{Phi_xi})   into (\ref{ber2}), then   equation (\ref{Psi}) into the resulting equation, we obtain a single equation  for the free surface  
\begin{equation}\label{bernoulli}
	\begin{split}
		-\ds\frac{c^2}{2}+\ds\frac{c^2}{2J}+\Y+\ds\frac{(\CC[(\Omega \Y+F)\Y_\xi])^2}{2J}-\ds\frac{\CC[(\Omega \Y+F)\Y_\xi]}{J}(c-(\Omega \Y+F)\X_\xi)\\-\ds\frac{(\Omega \Y+F)^2\Y_\xi^2}{2J}-\frac{c(\Omega \Y+F)\X_\xi}{J}+Fc+\Omega\left(\frac{\Omega \Y}{2}+F\right)\Y +\Omega M=B.
	\end{split}
\end{equation}
Observe that $\mathbf{X}_\xi=\ds 1-\CC_0[\mathbf{Y}_\xi]$ and   $J=\mathbf{X}_\xi^2+\mathbf{Y}_\xi^2$ are given in term of  $\Y(\xi)$. Consequently,    this is an equation whose unknowns are $\Y(\xi)$, $c$, $D$ and $B$. It is the aim of the next section to describe a numerical approach for computing solitary waves.

\subsection{Numerical method}\label{steadymethod}

Up to this point, we have transformed the free boundary problem (\ref{eq:travelling1})--(\ref{eq:travelling2})   into an nonlinear system of  two equations (  (\ref{depth}) and (\ref{bernoulli}) )  and four unknowns $\Y(\xi)$, $c$, $D$ and $B$. 	In order to get a system  that can  be handled by Newton's method, we add two extra equations.

We  fix the amplitude $A$ of the wave through 
\begin{equation}\label{amplitude}
	Y(0)-Y(-L/2)=A,
\end{equation}
and based on the limit (\ref{limite}) we impose that 
\begin{equation} \label{limite2}
	Y(-L/2)= 0.
\end{equation}

Consider a discrete version of  equations  (\ref{depth}), (\ref{bernoulli}), (\ref{amplitude}) and (\ref{limite2}) as follows. Let us take an evenly spaced grid in the $\xi$-axis in the canonical domain, say 
\begin{equation} \label{grid}
	\xi_j=-L/2+(j-1)\Delta\xi, \hs{10pt} j=1,...,N, \; \hs{10pt} \text{where } \Delta\xi=L/N, 
\end{equation}
with $N$ even. We impose symmetry about $\xi=0$ so that $Y_j=Y_{N-j+2}$, where $Y_j=\Y(\xi_j)$.
Fixing $\Omega$ and $F$, we have  $N/2+4$ unknowns: $Y_1, \cdots, Y_{N/2+1}$, $c$, $D$ and $B$. We satisfy 	 equation (\ref{bernoulli}) at the grid points (\ref{grid}). The  Fourier modes are computed by the Fast Fourier Transform (FFT) and derivatives in the $\xi$-variable are performed spectrally \cite{Trefethen}.   This yields a system with   $N/2+1$ equations $$\mathcal{G}_j (Y_1, \cdots, Y_{N/2+1}, c, D,B)  = 0 \quad j = 1, \cdots, N/2+1.$$  Equation (\ref{depth}) is discretized using the trapezoidal rule, which leads to the equation
\begin{equation*}
	\mathcal{G}_{N/2+2} (Y_1, \cdots, Y_{N/2+1}, c, D,B)= \frac{Y_1+Y_{N/2+1}}{2}+\sum_{j=2}^{N/2}Y_j+1-D=0.
\end{equation*}
Finally, we satisfy equations (\ref{amplitude}) and (\ref{limite2}), resulting in a system of the $N/2+4$ equations and $N/2+4$ unknowns, 
$$\mathcal{G}_{N/2+3} (Y_1, \cdots, Y_{N/2+1}, c, D,B)= Y_{N/2+1}-Y_1 - A = 0,$$
$$\mathcal{G}_{N/2+4} (Y_1, \cdots, Y_{N/2+1}, c, D,B)= Y_1 = 0.$$

The system is solved by Newton's method, where our initial guess is taken to be the well known solitary wave  solution for the classical (irrotational) Korteweg-de Vries equation, that is

\begin{align*}
	\Y(\xi)=A_0\sech^2\left(\sqrt{3A_0/4}\xi\right), \; c=1+\frac{A_0}{2}, 
\end{align*}
where $A_0$ is chosen small. From there, the idea is to make use of the continuation technique in $A$ and $\Omega$, where the prior converged solution is fed as initial guess to a new solution. The Jacobian matrix of the system is computed by finite difference and the stopping criterion  for the Newton's  method is 
$$ \frac{\sum_{j=1}^{N/2+4} \vert\mathcal{G}_j\vert}{N/2+4} < 10^{-10}.$$

In all experiments performed we used  $L = 1500$ -- which  it is important to make sure that the method indeed converges to a solitary wave solution. 

\section{Results} \label{Sec_Results}

In subsection \ref{section:free surface} we present some solitary waves computed through our numerical method. A comparison between such waves with a weakly nonlinear KdV equation is made in order to provide a validation of our numerical procedure.  Then, the main results of the paper are discussed in  subsection \ref{section:pressure}.
 

\subsection{Steady waves}\label{section:free surface}

Several numerical computations are available and provide a detailed characterization of the shape of the free surface wave in flows with constant vorticity.  More specifically, it is  known  that the crests of the waves become rounder as  $\Omega$ decrease. This has been shown   for periodic traveling waves \cite{Teles:1988,Vanden-Broeck96, Ko Europe, Ribeiro-Jr:2017,DyachenkoHur19b} and    for solitary waves \cite{Vanden-Broeck94}.

\begin{figure}[!h]
	\centering
	\includegraphics[scale=1]{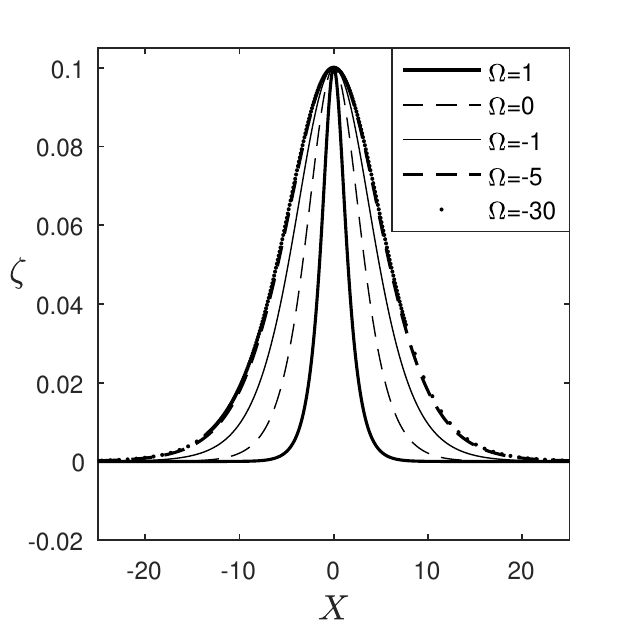}
	\includegraphics[scale=1]{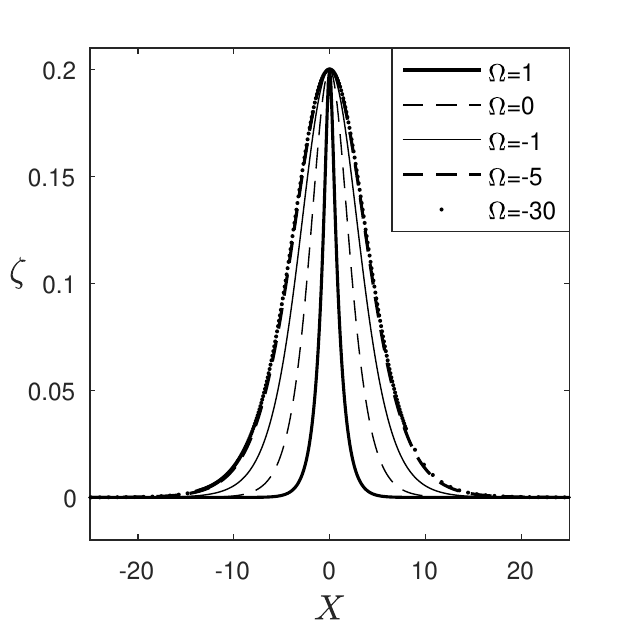}
	\caption{Wave profiles with amplitudes $A=0.1$ (left) and $A=0.2$ (right).}
	\label{profiles}
\end{figure}

Figure \ref{profiles} displays various wave profiles for different vorticity values. As can be seen, the numerical method captures these well known characteristics about waves with vorticity: more rounded or cuspidate profiles depending on the $\Omega$ sign. Although the computational domain used was  $L=1500$, for visualization purposes the plot window was chosen to be 50 units long. Besides, for each choice of $\Omega$, the Froude number  is fixed as  $F= {\Omega}/{2}$. This implies cancelling the average mass flow of the stream $\U_0=(\Omega Y + F,0)$. The choice of $F$ has no impact in the shape of wave nor in the location of the stagnation points and the appearance of pressure anomalies.  

\begin{figure}[!h]
	\centering
	\includegraphics[scale =1]{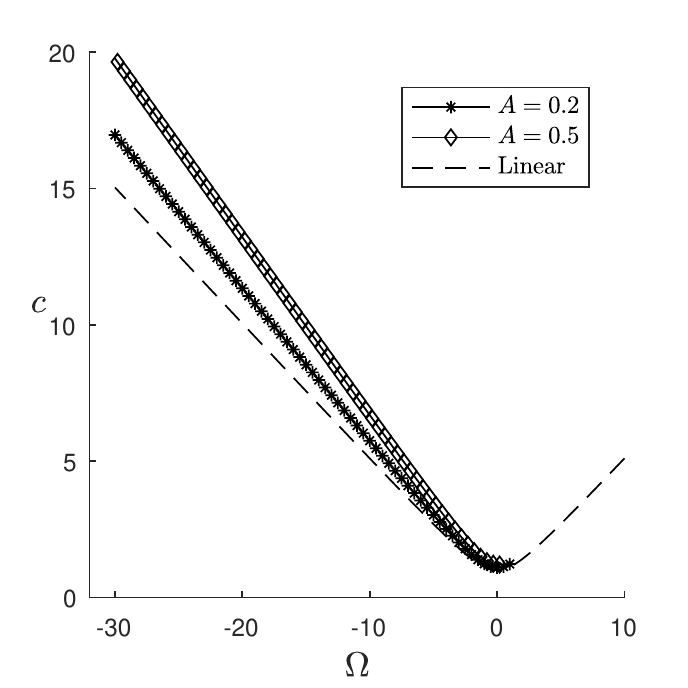}
	\caption{The wave speed as function of  $\Omega$ for different values of $A$. }
	\label{c_gamma}
\end{figure}

Furthermore, vorticity also has a straight-forward and expected effect in the velocity of the waves: greater vorticity implies greater velocity across the amplitude spectrum, a trend that matches with the well-known dispersion relation for  linear  long waves, as depicted in Figure \ref{c_gamma}. From that same figure, it is also notable that even though the method captures waves with negative $\Omega$ which are considerably big in modulus, convergence stops earlier in the positive direction. This phenomenon is in large part explained by the loss of solution regularity in a neighbourhood of $X=0$ when $\Omega$ becomes more positive, something hinted by Figure \ref{profiles}. The closed formula for the velocity shown in dashed lines in Figure \ref{c_gamma} is given by
\begin{equation*}
	c_{lin}=F-\ds\frac{\Omega}{2}+\sqrt{\ds\frac{\Omega^2}{4}+1}.
\end{equation*}

Beyond the linear theory, another model that can be used for comparison purposes is the weakly nonlinear   KdV equation. In what follows, we are interested  in investigating how the velocities are influenced by the increasing of amplitude for  a fixed vorticity. For small amplitudes, it is expected that waves computed should be similar to the $\sech^2$-type solution of the KdV equation. 

Regarding the analysis of the KdV model in the presence of vorticity we refer to the work of Guan \cite{Guan}.  The formulation presented by this author is used as benchmark of  our numerical solutions.

\begin{figure}[h]
	\centering
	\includegraphics[scale =1]{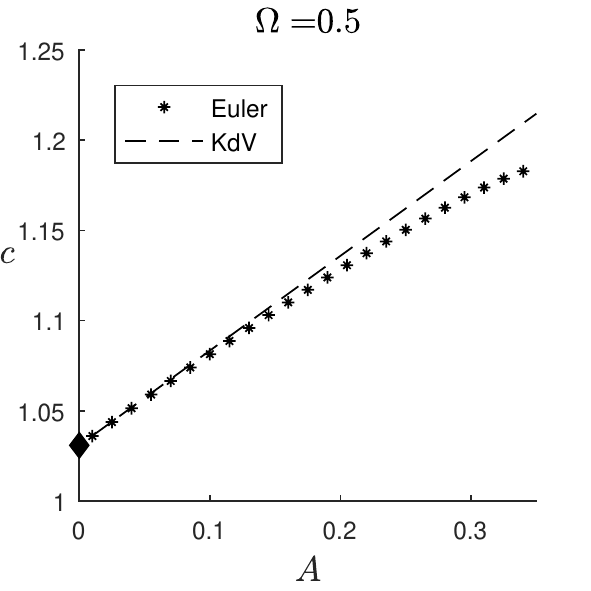}
	\includegraphics[scale =1]{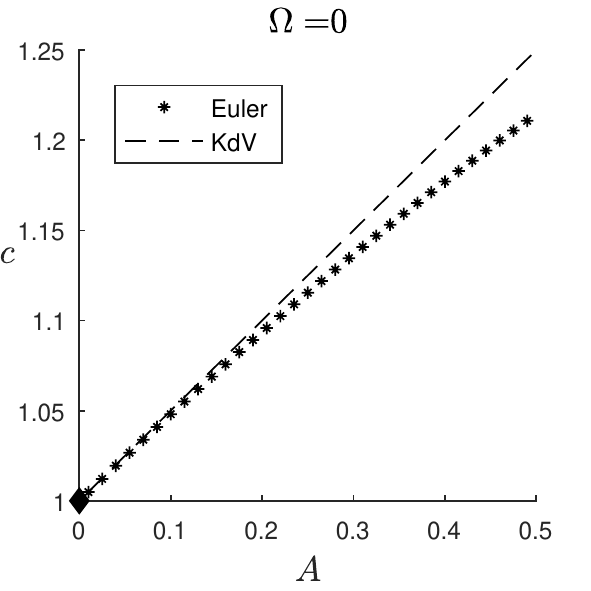}
	\includegraphics[scale =1]{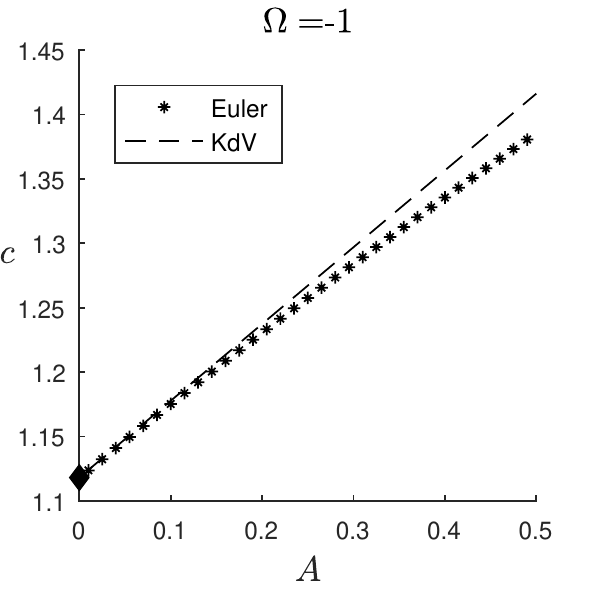}
	\includegraphics[scale =1]{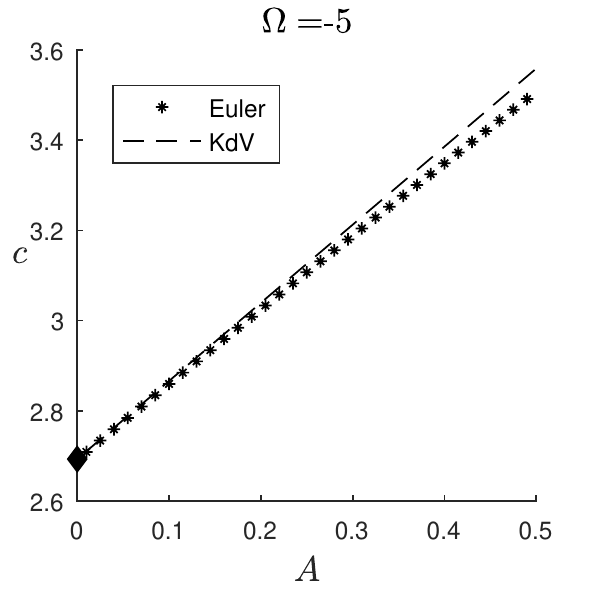}
	\caption{ The wave speed as function of the wave amplitude  for different choices of vorticity.}
	\label{KdV vel}
\end{figure}

For a given choice of parameters $\Omega$ and $A$, Figure \ref{KdV vel}  indicates the distance between our solutions to the analytical solution determined by the KdV equation. The dashed line displays the wave speed from the KdV solution after scaling to the Euler regime. As expected we see very close wave speed whenever $A$ is small but the overall pattern of speed/amplitude relation in the case of Euler solutions present a clear deviation from the linear distribution found in KdV. In particular, around $A=0.15$ and  $A=0.2$ we see a slight takeoff from the Euler regime in comparison to the KdV, while it is interesting to observe that the general aspect of this ``takeoff curve''  remains unchanged when we vary $\Omega$. For the interested reader, a study of the resolution of the numerical method is presented in Appendix \ref{reso}.

\subsection{Pressure in the bulk of the fluid} \label{section:pressure}

It is well known in the literature that pressure anomalies beneath nonlinear  periodic waves  are  connected to the arise of stagnation points \cite{Teles:1988,Vasan:2014,Ribeiro-Jr:2017}.  Starting from this point, we first investigate the appearance of stagnation points in terms of the intensity of the vorticity parameter ($\Omega$), then analyse the pressure within the bulk of the fluid.
For this purpose, we fix  solitary waves with amplitude $A = 0.2$,  $F = \Omega$  and let the vorticity vary. This choice of $F$ leads to a background flow  $(\Omega Y + F,0)$ with zero velocity at the bottom.

\begin{figure}[h!]
	\centering
	\includegraphics[scale=1]{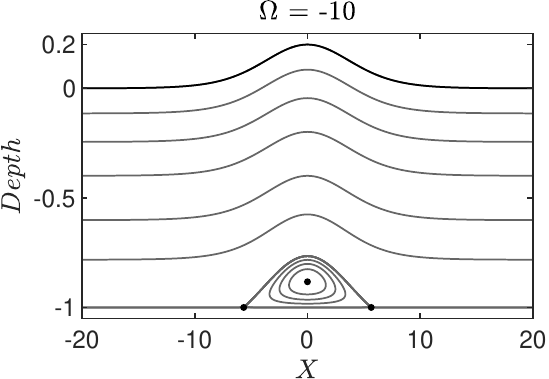} 
	\includegraphics[scale=1]{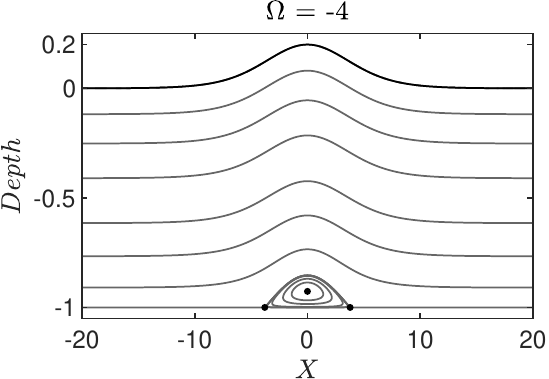} 
	\includegraphics[scale=1]{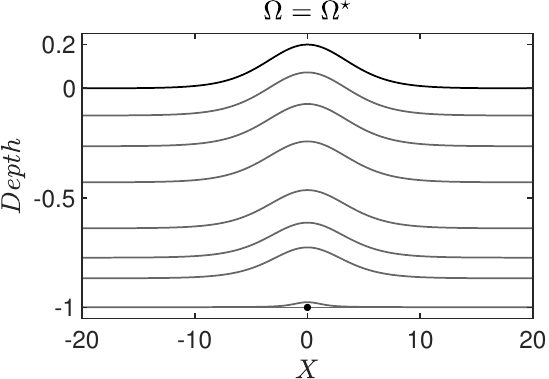} 
	\includegraphics[scale=1]{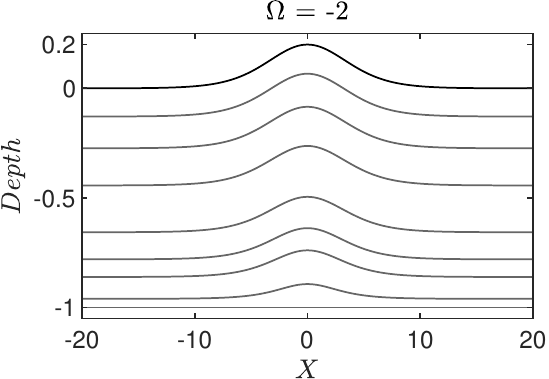} 
	\caption{Phase portraits  for different values of the vorticity parameter.  Circles correspond to  the location of the stagnation points. $\Omega^{\star}\approx -2.4967$.}
	\label{Fig1}
\end{figure}

Our first numerical  essay consists in  computing the phase portrait for different values of the vorticity parameter -- this is depicted in Figure  \ref{Fig1}. The markers represent the position of the stagnation points. 
We find that the stagnation points first appear on the bottom and below the crest for a critical value $\Omega^\star \approx -2.4967$. For $\Omega>\Omega^\star$ there is no stagnation points in the bulk of the fluid. Nonetheless, for $\Omega<\Omega^\star$   we obtain a flow with three stagnation points: two saddles located at the bottom and one centre located within the bulk of the fluid and below the crest -- forming a region with closed streamlines which is described as a single   Kelvin cat's eye structure.   As the vorticity gets stronger this structure becomes wider, i.e, the saddles remain on the bottom moving  away from each other and the centre moves upwards.

\begin{figure}[h!]
	\centering
	\includegraphics[scale=1.2]{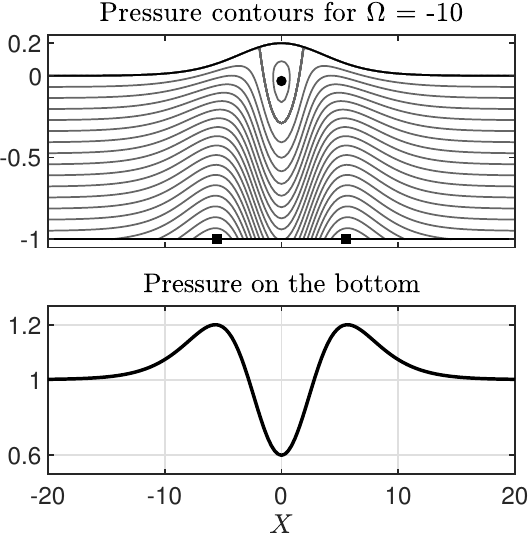} 
	\includegraphics[scale=1.2]{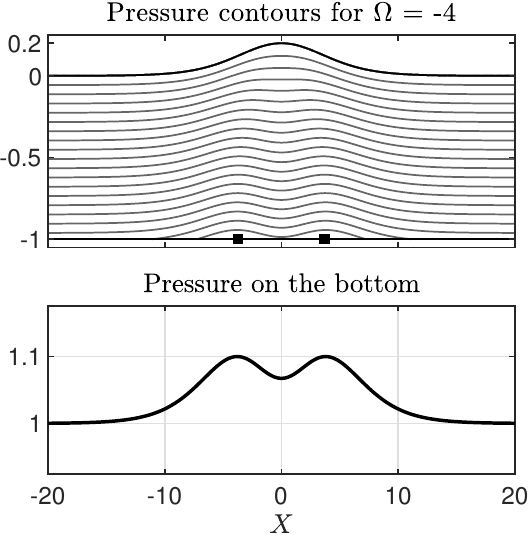} 
	\includegraphics[scale=1.2]{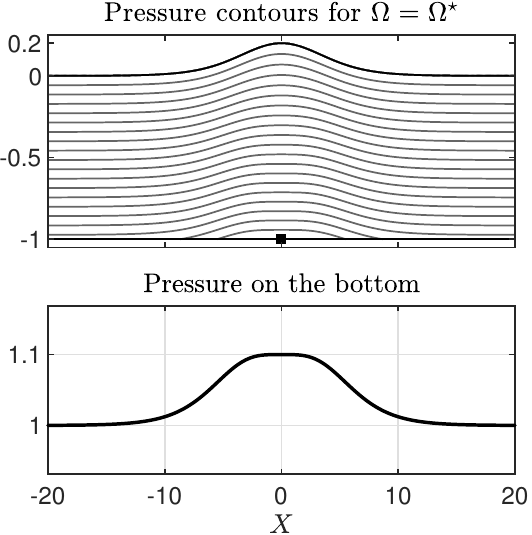} 
	\includegraphics[scale=1.2]{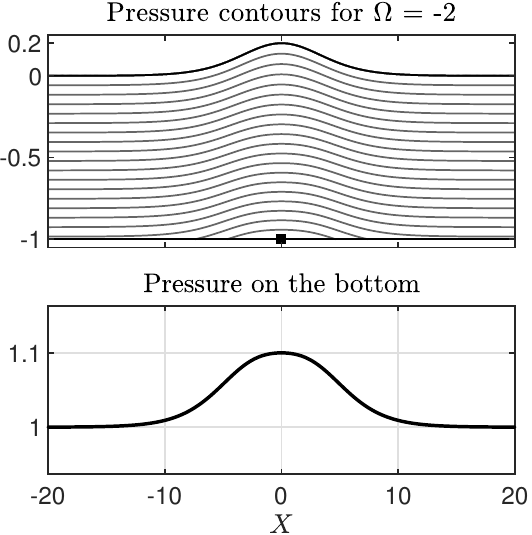} 
	\caption{ Pressure beneath the solitary wave with amplitude $ A = 0.2$ and its correspondent pressure on the bottom boundary.  Circle and square markers indicate the location of global minima and maxima of the pressure respectively.}
	\label{Fig2}
\end{figure}

Figure \ref{Fig2} shows the pressure contours and the pressure on the bottom boundary for the same waves depicted in Figure \ref{Fig1}. We notice that according to the value of $\Omega$ the following anomalies occur: (i) the maximum pressure value may not  be attained at the bottom and below the crest; (ii) the minimum pressure value may be attained within the bulk of the fluid; (iii) the pressure on the bottom boundary may have two local maxima. 
These anomalies have been observed  for nonlinear periodic waves \cite{Teles:1988,Vasan:2014, Ribeiro-Jr:2017} and for overhanging solitary waves \cite{Strauss:2016}, but to the best of our knowledge this the first time that such results are reported for solitary waves that are graph of a function.  These results indicate that the anomalies in the pressure and the stagnation points are somehow related. Moreover, pressure anomalies and stagnation points occur  when  a threshold vorticity is achieved. However,  a detailed  theoretical study is necessary to explain such phenomenon.

\section{Conclusion}

In the present work, we have studied the pressure beneath solitary waves in flows with constant vorticity. Our results indicate that there exists a threshold vorticity such that pressure anomalies and stagnation points  occur when the intensity of the  vorticity  is greater than the threshold. More specifically, when the vorticity is below  this threshold the pressure on the bottom boundary has one local maximum and there is no stagnation point in the flow. Once the vorticity crosses this threshold the pressure on the bottom boundary has two local maxima and the flow has three stagnation points (one centre and two saddles).

\section*{Acknowledgements}

The author E.M.C. is grateful for the financial support provided by CAPES Foundation (Coordination for the Improvement of Higher Education Personnel) during part of the development of this work.

\bibliographystyle{abbrv}

\appendix 

\section{Resolution study}\label{reso}
In what follows we show that the method is  independent of the grid size by calculating the distance between outputs for different choices of $\Delta \xi$. These experiments were performed for waves with amplitude $A=0.2$.  We take the reference grid  as  $\Delta \xi^*=0.0458$, the  finest resolution computed.



\begin{table}[h]
	\begin{center}
			\begin{tabular}{r|c|r|r}
				\hline
				$\Omega$ & $\Delta\xi$ & $\dfrac{\|\zeta_{\Delta \xi} - \zeta^*\|_2}{\|\zeta^* \|_2}$ & $\dfrac{\vert c_{\Delta \xi} - c^*\vert}{\vert c^* \vert}$ \\
				\hline
				\multirow{4}{*}{$0$}    & $0.0916$ & $1.6\times10^{-10}$ & $2.5\times10^{-12}$ \\   
				& $0.1831$ & $4.6\times10^{-10}$ & $7.4\times10^{-12}$ \\  
				& $0.3662$ & $6.7\times10^{-8}$ & $5.6\times10^{-10}$ \\ 
				& $0.7324$ & $1.2\times10^{-4}$ & $5.2\times10^{-6}$ \\ 
				\hline
				\multirow{4}{*}{$-1$}     & $0.0916$ & $5.4\times10^{-13}$ & $2.1\times10^{-14}$ \\ 
				& $0.1831$ & $5.7\times10^{-13}$ & $1.0\times10^{-14}$ \\  
				& $0.3662$ & $1.5\times10^{-12}$ & $1.3\times10^{-14}$ \\  
				& $0.7324$ & $1.2\times10^{-7}$ & $1.8\times10^{-9}$ \\ 
				\hline
				\multirow{4}{*}{$1$}    & $0.0916$ & $6.8\times10^{-5}$ & $1.0\times10^{-6}$ \\   
				& $0.1831$ & $0.0027$ & $1.7\times10^{-4}$ \\ 
				& $0.3662$ & $0.0477$ & $0.0015$ \\  
				& $0.7324$ & $0.1164$ & $0.0102$  \\ 
				\hline
			\end{tabular} 	\label{tabela2}
	\end{center}
			\caption{Resolution study for waves of amplitude $A=0.2$}\label{tab1}%
\end{table}

 In table \ref{tabela2}, we denote by $\zeta_{\Delta \xi}$  the wave profile and by $c_{\Delta \xi}$    the wave speed  obtained from the Newton's method using a grid  with size $\Delta \xi$. In addition, we consider as  $\zeta^*$  
and  $c^*$ the wave profile and its speed computed in the finest grid. 
These experiments were performed for waves with amplitude $A = 0.2$. 
Note that for  $\Omega = 1$ the numerical scheme requires more resolution for approximating  the solution with more accuracy. This  can be explained by a combination of two factors: i) the emergence of cusps; ii) the issue of crowding phenomenon present in conformal mappings. For this reason, finer grids are necessary to accurately  compute waves in presence of currents where  $\Omega$ is positive.

\end{document}